\documentclass[letter]{aa} % for the letters 
\usepackage{graphicx}
\usepackage{txfonts}
\usepackage{hyperref}  
\hypersetup{colorlinks=true,linkcolor=[rgb]{1.,0.2,0.2},citecolor=[rgb]{0.1,0.4,1.},filecolor=[rgb]{0.7,0.2,0.2},urlcolor=[rgb]{0.7,0.2,0.2}}
\usepackage{color}
\definecolor{blue}{rgb}{0., 0., 1}
\newcommand {\T}{Table\,}

\newcommand {\Fig}{Fig.\,}

\begin{document}

   \title{Abell 0399$-$Abell 0401 radio bridge spectral index: \\ the first multifrequency detection}

   \author{G.V. Pignataro
          \inst{1,2} \fnmsep\thanks{e-mail: \href{mailto:giada.pignataro2@unibo.it}{giada.pignataro2@unibo.it}}
          \and
          A. Bonafede\inst{1,2}
          \and 
          G. Bernardi\inst{2,3,4}
          \and
          F. de Gasperin\inst{1,5}
          \and
          G. Brunetti\inst{2}
          \and
          T. Pasini\inst{2,5}
          \and
          F. Vazza\inst{1,2,5}
          \and
          N. Biava\inst{2,7}
          \and
          J. M. G. H. J. de Jong\inst{6}
          \and
          R. Cassano\inst{2}
          \and
          A. Botteon\inst{2,1}
          \and
          M. Br\"{u}ggen\inst{5}
          \and
          H. J. A. R\"{o}ttgering\inst{6}
          \and
          R. J. van Weeren\inst{6}
          \and
          T. W. Shimwell\inst{6,8}
          }

   \institute{Dipartimento di Fisica e Astronomia, Universit\`a degli Studi di Bologna, via P. Gobetti 93/2, 40129 Bologna, Italy
        \and
             INAF -- Istituto di Radioastronomia, via P. Gobetti 101, 40129 Bologna, Italy
        \and
            Department of Physics and Electronics, Rhodes University, PO Box 94, Makhanda, 6140, South Africa
        \and
            South African Radio Astronomy Observatory (SARAO), Black River Park, 2 Fir Street, Observatory, Cape Town, 7925, South Africa
        \and
            Hamburger Sternwarte, Universitat Hamburg, Gojenbergsweg 112, D-21029, Hamburg, Germany 
        \and    
            Leiden Observatory, Leiden University, PO Box 9513, 2300 RA Leiden, The Netherlands
        \and
             Thüringer Landessternwarte, Sternwarte 5, D-07778 Tautenburg, Germany
        \and
            ASTRON, the Netherlands Institute for Radio Astronomy, Postbus 2, 7990 AA Dwingeloo, The Netherlands}

   \date{Received September 15, 1996; accepted March 16, 1997}

% \abstract{}{}{}{}{} 
% 5 {} token are mandatory
 
  \abstract
  % context heading (optional)
  % {} leave it empty if necessary  
   {}
  % aims heading (mandatory)
   {Recent low-frequency radio observations at 140~MHz discovered a 3~Mpc-long bridge of diffuse emission connecting the galaxy clusters Abell 0399 and Abell 0401. We present follow-up observations at 60~MHz to constrain the spectral index of the bridge, which so far has only been detected at 140 and 144~MHz.}
  % methods heading (mandatory)
   {We analysed deep ($\sim$18 hours) LOw Frequency ARray (LOFAR) Low Band Antenna (LBA) data at 60~MHz to detect the bridge at very low frequencies. We then conducted a multi-frequency study with LOFAR HBA data at 144~MHz and uGMRT data at 400~MHz. Assuming second-order Fermi mechanisms for the re-acceleration of relativistic electrons driven by turbulence in the radio bridge regions, we compare the observed radio spectrum with theoretical synchrotron models.}
  % results heading (mandatory)
   {The bridge is detected in the $75''$ resolution LOFAR image at 60~MHz and its emission fully connects the region between the two galaxy clusters. Between 60~MHz and 144~MHz we found an integrated spectral index value of $\alpha_{60}^{144}= -1.44 \pm 0.16$ for the bridge emission. For the first time, we produced spectral index and related uncertainties maps for a radio bridge. We produce a radio spectrum, which show significant steepening between 144 and 400~MHz.}
  % conclusions heading (optional), leave it empty if necessary 
   {This detection at low frequencies provides important information on the models of particle acceleration and magnetic field structure on very extended scales. The spectral index gives important clues to the origin of inter-cluster diffuse emission. The steepening of the spectrum above 144~MHz can be explained in a turbulent re-acceleration framework, assuming that the acceleration timescales are longer than $\sim$ 200~Myr.}


   \maketitle
%
%________________________________________________________________

\section{Introduction}

Matter accretes primarily onto galaxy clusters along filaments of the so-called cosmic-web, and the subsequent merger of these systems releases an extreme amount of energy into the Intra-cluster medium (ICM) \citep{markevitch2007}. Radio observations have provided strong evidence for the processes where this energy is channeled into particle acceleration and magnetic field amplification in the form of diffuse emission with steep synchrotron spectra ($\alpha<-1$, with flux density $S_{\nu}\propto \nu^{\alpha}$). In galaxy clusters, we observe different types of radio diffuse sources: radio relics, mini halos and radio halos \citep[for an extensive review, see][]{vanweeren2019}. Recent studies unveiled the existence of diffuse radio emission on an even larger scale \citep{govoni2019,botteon2020,hoeft2021}. This emission extends beyond the cluster centers and traces the densest region of cosmic-web filaments, where the gas is compressed during the first phase of merger between clusters. Multifrequency studies of synchrotron emission from radio-bridges between clusters can shed light on mechanisms of particle acceleration and properties of the magnetic fields on poorly probed scales \citep{vazza2019}.
\begin{figure*}[h!]
   \centering
   \includegraphics[width=1\linewidth]{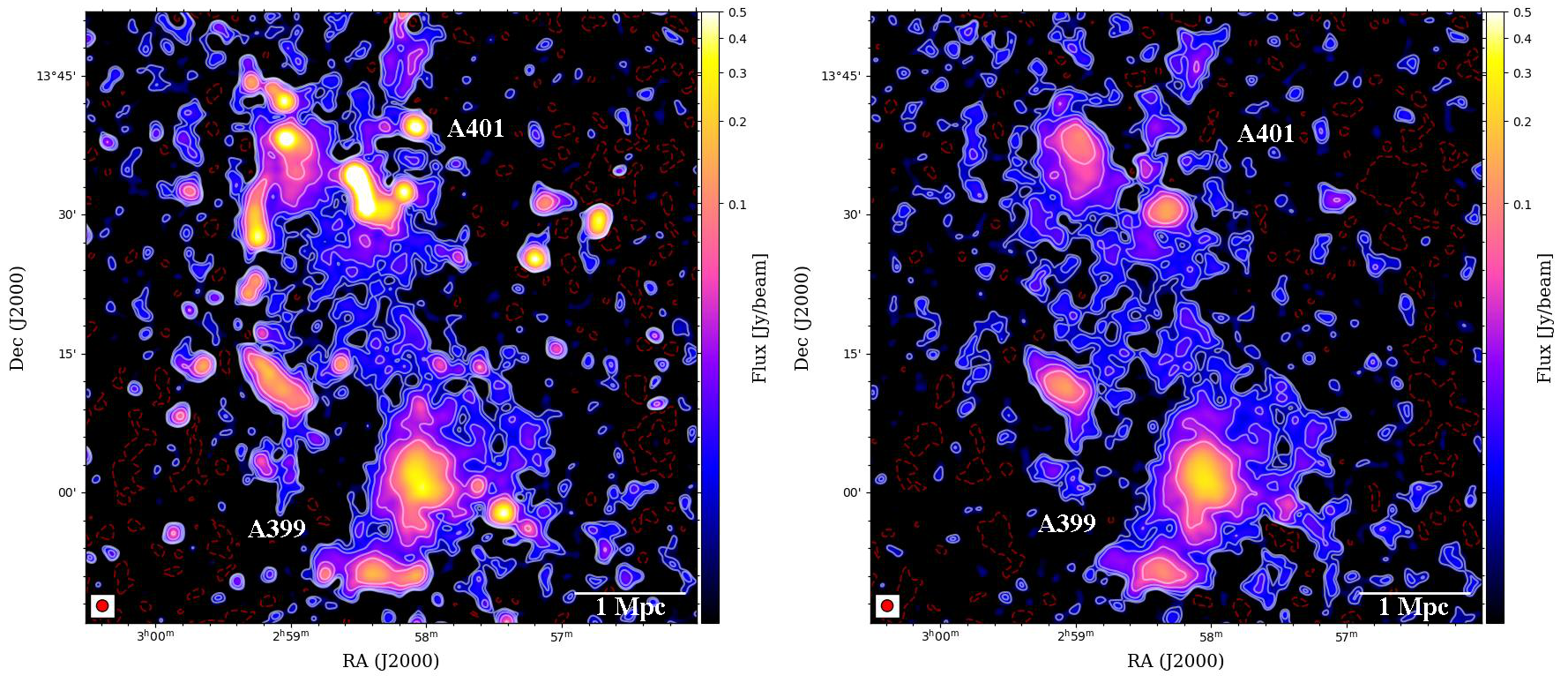}
   \caption{Radio images at 60~MHz of the A399-A401 clusters pair. \textit{Left panel:} Low-resolution ($75^{\prime\prime}$, beam-size in bottom-left corner) image with compact sources and $\sigma_{\rm rms}= 3$mJy~beam$^{-1}$ produced with the parameters listed in \T\ref{tab:imaging}. \textit{Right panel:} Same as left panel, but after compact source subtraction. Contour levels (in white) start at $2\sigma_{\rm rms}$, $3\sigma_{\rm rms}$, $5\sigma_{\rm rms}$, and then increase to $20\sigma_{\rm rms}$ with factors of 2. A negative contour at $-3\sigma_{\rm rms}$ is shown in red.}
              \label{images}%
\end{figure*}
The galaxy clusters Abell 0399 and Abell 0401 (hereafter, A399 and A401) are a local ($z\sim0.07$, \citealt{oegerle2001}) pair found in a pre-merger state \citep{bonjean2018} where X-ray observations \citep{fujita1996, fujita2018, akamatsu2017} revealed the presence of $6-7$~keV ionised plasma in the region between the clusters. The gas in this region is also detected via the Sunyaev-Zeldovich (SZ) effect by \textit{Planck} \citep{planck2013,planck2016,bonjean2018} and the Atacama Cosmology Telescope (ATC) \citep{hincks2022, radiconi2022}. The first radio bridge connecting two galaxy clusters is discovered in this system \citep{govoni2019}. The bridge is detected at 140~MHz with the LOw Frequency ARray (LOFAR), and extends for approximately $3$~Mpc, which is the projected separation of the two clusters that also host radio halos.
Following the discovery of the bridge, the A399-A401 system was extensively studied at radio frequencies. \cite{dejong2022} presented a 40-hour, deep LOFAR observation at 144~MHz and investigated further the properties of the diffuse emission in the bridge. They were able to detect the bridge at high significance, and measure a flux density of $S_{144}=550\pm60\text{ mJy}$ over $2.7\text{ Mpc}^{2}$.
To cause diffuse emission on such scales, electrons would need to be generated or re-accelerated in situ because of their short synchrotron life-times. \cite{govoni2019} proposed the model of Fermi-I re-acceleration of fossil electrons by weak shocks crossing the region, which would result in spectral indices similar to those of radio relics ($\alpha\sim-1.3)$. Alternatively, \cite{brunetti2020} showed how this emission could also be explained by a Fermi-II re-acceleration process. In this scenario, the fossil relativistic particles are re-accelerated by turbulence in amplified magnetic fields over Mpc-scales. This would result in steep observed synchrotron spectra between $150$~MHz and $1.5$~GHz ($\alpha<-1.5$). 
Recently, \cite{nunhokee2021} presented WSRT observations at 346 MHz that were not sufficiently deep to observe the bridge, and therefore place a limit on the bridge spectral index ($\alpha^{346}_{140}<-1.5$). A similar procedure to place limits on the emission of radio bridges is defined in \cite{pignataro2023}. The non-detection of bridge emission in high sensitivity uGMRT data at 400 MHz results in a more stringent constraint on the steep bridge spectral index ($\alpha^{400}_{140}<-2.2$), disfavouring the Fermi-I acceleration scenario.\\

Other than the detection in A399-A401, only a few other radio bridges associated with merging clusters are known. \cite{botteon2018,botteon2020} report the bridge in Abell 1758N-S where they are also able to measure a spectral index for a patch of emission. Moreover, a candidate bridge is reported in Abell 1430 \citep{hoeft2021}, and the bridge between relic and halo in A1550 \citep{pasini2022}. Recently, a few more bridges between clusters and groups have been discovered \citep[see][]{bonafede2021,venturi2022}. However, for none of these objects has it been possible to determine the spectral index of the extended diffuse emission.\\

In this Letter, we present a multifrequency study conducted with new LOFAR Low Band Antenna (LBA) data at 60~MHz that allows, for the first time, the determination of the spectral index of the radio bridge in the A399-A401 bridge between 60 and 144~MHz. Here, we assume a $\Lambda$CDM cosmology, with $H_{0}=70$ km s$^{-1}$~Mpc$^{-1}$, $\Omega_{m}=0.3$, and $\Omega_{\Lambda}=0.7$. With these assumptions, at the average distance of the A399-A401 system (z$\sim0.07$), $1^{\prime}=83$~kpc and the luminosity distance is $D_{\rm L}= 329$~Mpc.

%__________________________________________________________________

\section{Observations and data reduction}
\begin{figure}[h!]
   \centering
   \includegraphics[width=1\linewidth]{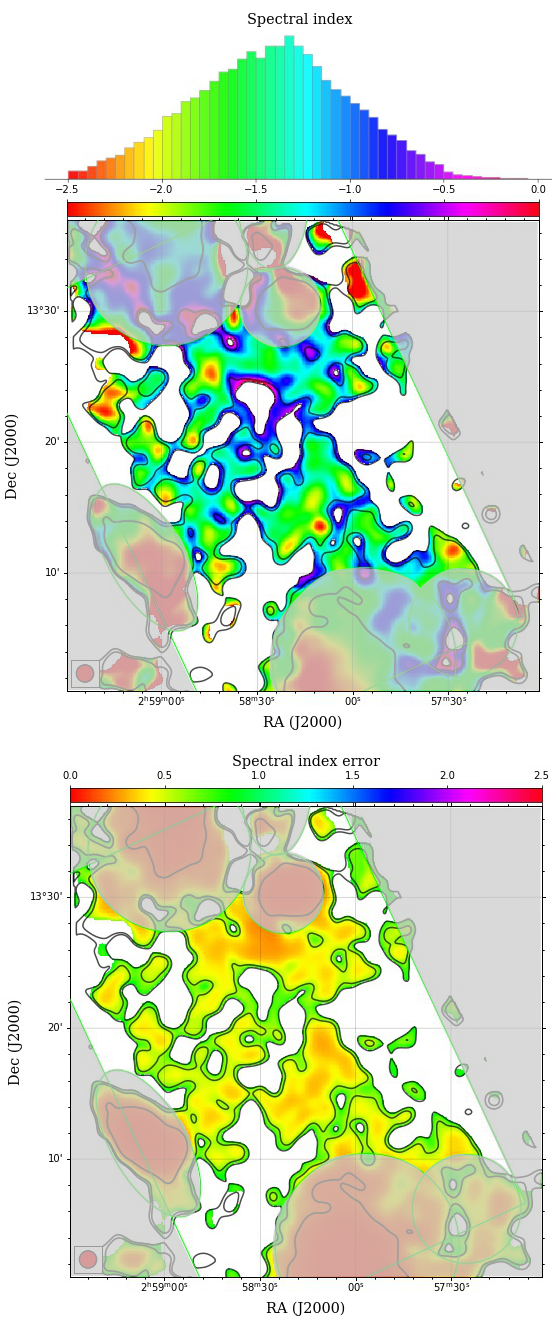}
      \caption{Spectral index maps. \textit{Top panel:} Spectral index map of the emission in the radio bridge in A399-A401 between 60 and 144 MHz, with resolution $80''$. The spectral index values distribution (histogram on top) only refers to the bridge emission inside the $2\sigma_{\rm rms}$ level, not covered by the grey mask over the radio halos. \textit{Bottom panel:} Associated spectral index error map. Overlaid in black are the LOFAR LBA contours at $2\sigma_{\rm rms}$ and $3\sigma_{\rm rms}$.
              }
         \label{fig:spixmaps}
   \end{figure}

The observations of A399-A401 are 18 hours long and were carried out using the LOFAR LBA system in the frequency range of $30-77$ MHz (proposal code: LC$13\_022$, P.I.: Bonafede). 
The correlated data are initially pre-processed for radio frequency interference (RFI) and Demix procedure \citep{vandertol2009, degasperin2020a}. In this case, both Cygnus A and Cassiopia A were demixed. The data are also averaged from $1$ to $4$ s integration time and from $64$ to $8$ channels per subband to reduce memory and computational load.
The data are then calibrated with the Library for Low Frequencies (LiLF) pipeline. Here we briefly summarize its main steps here, while a full description can be found in \cite{degasperin2019, degasperin2020}. The pipeline obtains solutions from the calibrator (3C196) and applies them to correct the target data. In this part, the pipeline isolates well the systematic effects of the polarisation alignment, the bandpass, and Faraday rotation. The clock drift is left mixed with the total electron content (TEC), and both are transferred to the target.  Additional calibration is then needed to correct for the differential ionospheric effects that strongly affect the lower frequencies \citep{degasperin2018}, especially for observations at low elevation, as it is the case of this target (Dec  +$13^{\circ}$). This is done for the target field with direction-independent (DI) self-calibration, and then direction-dependent (DD) calibration. After a round of calibration, we inspected the data and found very strong artefacts produced by a radio galaxy (3C79) outside the first primary beam null. To mitigate this effect, we reduced the observation bandwidth to eliminate the frequencies where the source is the brightest and the primary beam is largest. This reduced the frequency range to $44-77$~MHz, with the central frequency at $60$~MHz. 
Then, we proceeded with a new round of DI phase and amplitude self-calibration, which performs two cycles and corrects for the systematic errors in the target field. 
From the sky-model produced in the last round of DI calibration, the bright sources are selected as calibrators for the DD calibration, which removes the differential ionospheric errors in the direction of each calibrator within the field of view. Both steps are described in detail in \cite{degasperin2019}. 
Finally, the pipeline performs the target extraction \citep{vanweeren2021, biava2021, pasini2022}, where the direction-dependent solutions found are used to subtract all sources outside a radius of approximately 0.5° around the target system. To refine the calibration towards the target, a few cycles of phase self-calibration at increasing time-resolution are performed on the extracted field. At this point, we can use the final calibrated extracted visibilities of the target to image at different resolutions with \texttt{WSClean v3.1} \citep{offringa2014}.\\
We produced a final primary beam corrected image at the central frequency of 60 MHz, at a resolution of $75''$ with a rms noise of $\sigma_{\rm rms}=3$ mJy~beam$^{-1}$, shown in \Fig\ref{images} (left panel). We then produced a high-resolution image excluding baselines shorter than $900\lambda$ (i.e. emission on scales more extended than $\sim4'$) to recover only the compact sources, and then subtracted their components from the visibilities. The $75''$ resolution source-subtracted image of the target field is shown in \Fig\ref{images} (right panel). The radio bridge is detected at $2\sigma_{\rm rms}$ connecting continuously the two radio halos, and shows a patchier morphology at $3\sigma_{\rm rms}$ level.

% Two column figure (place early!)

%______________________________________________ Gamma_1 (lg rho, lg e)

%
% Please add the following required packages to your document preamble:
% \usepackage{graphicx}

\begin{table}[]
\tiny
\renewcommand{\arraystretch}{1.2}

\centering
\resizebox{\columnwidth}{!}{%
\begin{tabular}{lcc}
                                         & \textbf{LBA}                            & \textbf{HBA\textsuperscript{*}}      \\ \hline \hline
                                    
\multicolumn{1}{l|}{Image size (pixels)}     & \multicolumn{1}{c|}{1500}               & 1500               \\
\multicolumn{1}{l|}{Cell size (arcsec)}      & \multicolumn{1}{c|}{6}                  & 6                  \\
\multicolumn{1}{l|}{Weighting}           & \multicolumn{1}{c|}{Briggs robust -0.5} & Briggs robust -0.5 \\
\multicolumn{1}{l|}{min-uv ($\lambda$)}  & \multicolumn{1}{c|}{24}                 & 24                 \\
\multicolumn{1}{l|}{max-uv ($\lambda$)}  & \multicolumn{1}{c|}{3500}               & -                  \\
\multicolumn{1}{l|}{Taper gaussian (arcsec)} & \multicolumn{1}{c|}{70}                 & 60                 \\ 
\multicolumn{1}{l|}{$\sigma_{\rm rms}$ (mJy~beam$^{-1}$)} & \multicolumn{1}{c|}{3}                 & 0.5                 \\ \hline

\end{tabular}%
}
\smallskip
\caption{\texttt{WSClean} imaging parameters used to produce the low-resolution source-subtracted images for spectral index analysis. In the last line, we report the image rms noise $\sigma_{\rm rms}$. \textsuperscript{*}The HBA low-resolution images at 144 MHz made with these parameters are presented in \cite{dejong2022}.}
\label{tab:imaging}
\end{table}
%______________________________________________________________

\section{Results and discussion}
\subsection{Spectral analysis}

In order to understand the origin and properties of the large-scale emission we investigate the integrated spectral index and spectral index distribution of the bridge emission. 
To perform the spectral index analysis, the LBA data are imaged with the same parameters as the HBA data at 144 MHz presented in \cite{dejong2022}. The imaging parameters are listed in \T\ref{tab:imaging}. In particular, we matched the \textit{uv-}min and weighting scheme, to recover the same angular scales and reach a similar resolution between LBA and HBA observations. To ensure we are correcting for possible shifts introduced by the phase self-calibration, we checked that the locations of the peaks of some point-sources in the field are matching in both images. We then convolve the images to the same restoring beam ($80''$ resolution). 
Additionally, we performed a flux density alignment on the uv-subtracted image HBA maps presented in \cite{dejong2022}, as usually done for LoTSS pointings \citep{lotssdr1, lotssdr2}, and applied a scale factor of $0.9$ to the data.
Finally, we considered only the emission above $2\sigma_{\rm rms}$ contour in both images and computed the spectral index map with associated 
error map, assuming a flux calibration error of $10\%$ \citep[as done for LoLSS,][]{degasperin2021}.
   
We show the resulting spectral index and spectral index error maps between 60 and 144 MHz with a resolution of $80"$ in \Fig\ref{fig:spixmaps}. We consider only the emission outside the grey mask as part of the radio bridge, while we mask the radio halos and other features of diffuse emission not related to the bridge emission. \Fig\ref{fig:spixmaps} (top panel) shows the distribution of spectral index along the bridge, and the occurrence of each value is represented in the histogram. The distribution appears overall uniform, with most values found between $-1.5\leq\alpha\leq-1.2$ . The error map (\Fig\ref{fig:spixmaps}, bottom panel) shows the associated errors, that are mostly around $\Delta\alpha\sim0.2$.

Within the $2\sigma_{\rm rms}$ level contours of the LBA image we measure a flux density of $S_{\rm 60~MHz} = 1.77 \pm 0.18$ Jy and $S_{\rm 144~MHz} = 0.50 \pm 0.05$ Jy\footnote{This is in agreement with the flux density measured in \cite{dejong2022}, the difference is due only to different areas.}, leading to an integrated spectral index value for the radio bridge of $\alpha_{60}^{144}= -1.44 \pm 0.16$. This is the first estimate of a radio bridge spectral index, and it provides important information on the models of magnetic field amplification and particle re-acceleration processes on megaparsec-scales. 

\subsection{Theoretical models}\label{models}

The origin of the radio emission from radio bridges is still being investigated. The Mpc-scale size of the bridge requires an in-situ mechanism to accelerate the relativistic particles to travel over these scales \citep{brunetti2014}. \cite{govoni2019} 
suggested a shock-driven emission model, where multiple weak shocks re-accelerates a pre-existing population of electrons. However, they show that it is difficult to account for the extension and strength of the bridge emission only via shock prior to the collision between A399 and A401. Moreover, the high-sensitivity study in \cite{dejong2022} reported that they do not observe filamentary structures or shock surfaces in the bridge region, disfavouring the shock origin.
We want to investigate the spectrum of the bridge emission with the measured flux densities at LOFAR frequencies. Additionally, we want to incorporate data from the uGMRT observations at a central frequency of 400~MHz presented in \cite{pignataro2023}, where the bridge emission is undetected. This also allows for a comprehensive comparison of the radio spectrum with the synchrotron spectrum predicted by theoretical models.
In \cite{pignataro2023} we found a limit on the bridge emission by following the injection procedure. In order to incorporate the limit in the  radio spectrum produced in this work, we need to perform again the procedure injecting the 60~MHz emission model, where the emission appears less extended in than the 144~MHz detection. 

Following an extended approach to the injection procedure with the 60~MHz observation as a starting model, we find a limit for the spectral index between 60~MHz and 400~MHz of $\alpha^{400}_{60}<-1.75$ at a 95\% confidence level. The injection method is discussed in Appendix \ref{appendix}.
\begin{table}[]
%{\fontsize{4}{5.5}\selectfont 
\renewcommand{\arraystretch}{1.2}

\centering
\resizebox{\columnwidth}{!}{%
\begin{tabular}{cccc}

{\tiny \textbf{Telescope}} &{\tiny $\mathbf{\nu}$ \textbf{{ [}MHz{]}}}  & {\tiny $\mathbf{P(\nu)}$\textbf{{ [}W Hz$^{-1}${]}}}                  & {\tiny$\mathbf{S(\nu)}$\textbf{{ [}Jy{]}}} \\ \hline \hline 
{\tiny LOFAR LBA} & \multicolumn{1}{|c|}{{\tiny 60}}  & \multicolumn{1}{c|}{{\tiny$(2.2 \pm 0.2) \times 10^{25}$}} & {\tiny$1.77 \pm 0.18 $}          \\
{\tiny LOFAR HBA} & \multicolumn{1}{|c|}{\tiny 144} & \multicolumn{1}{c|}{\tiny $(6.1 \pm 0.6) \times10^{24}$}  & {\tiny$0.50 \pm 0.05 $}          \\
{\tiny uGMRT Band3} & \multicolumn{1}{|c|}{\tiny 400} & \multicolumn{1}{c|}{\tiny $< 7.6 \times 10^{23}$}         & {\tiny $<0.06$}                   \\ \hline
                        
\end{tabular}%
}
\smallskip
\caption{Radio quantities for the bridge emission at the three frequencies presented in this study. Columns: (1) Telescope; (2) Central frequency in MHz; (3) Radio luminosity in W Hz$^{-1}$ at given frequency; (4) Integrated flux density in Jy at given frequency. }
\label{tab:flux_values}
\end{table}

\begin{figure}[h!]
   \centering
   \includegraphics[width=1\linewidth]{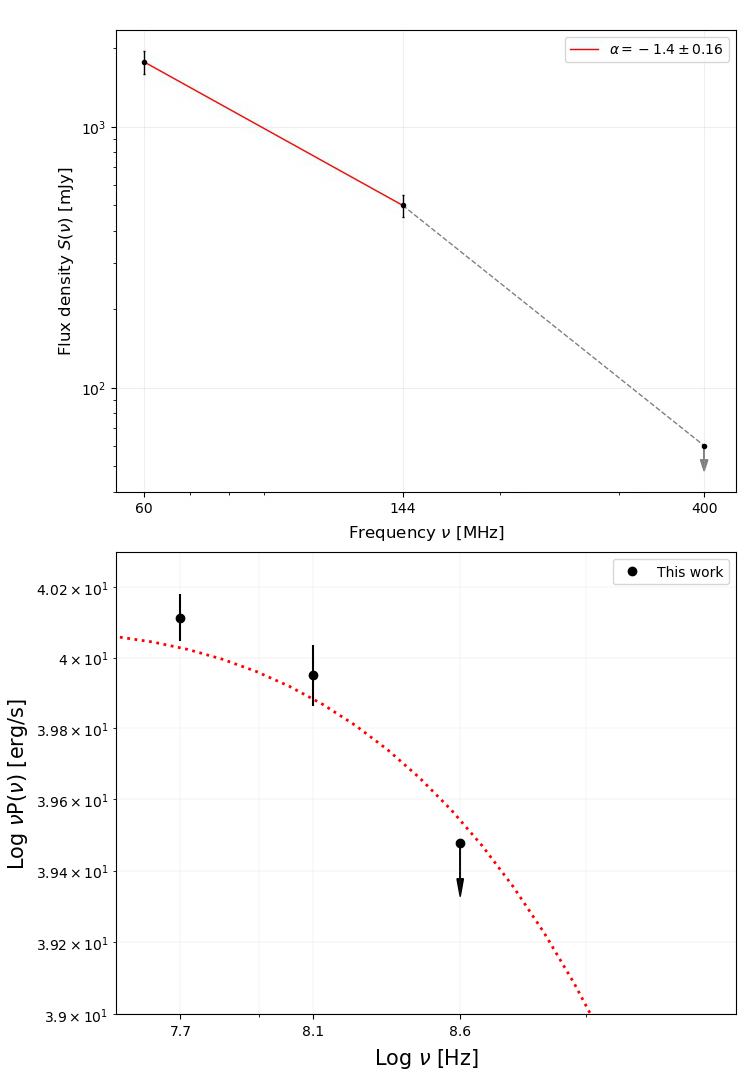}
      \caption{Radio spectra. \textit{Top panel:} The bridge emission radio spectra in integrated flux density. The grey arrow represents the uGMRT limit. \textit{Bottom panel:}  Radio luminosity of the bridge emission compared to a synchrotron theoretical model (red curve) produced by the relativistic particle populations with acceleration times $\tau_{\rm acc}>$200~Myr. 
              }
         \label{fig:spectrum}
   \end{figure} 

   \begin{figure*}[h!]
   \centering
   \includegraphics[width=1\linewidth]{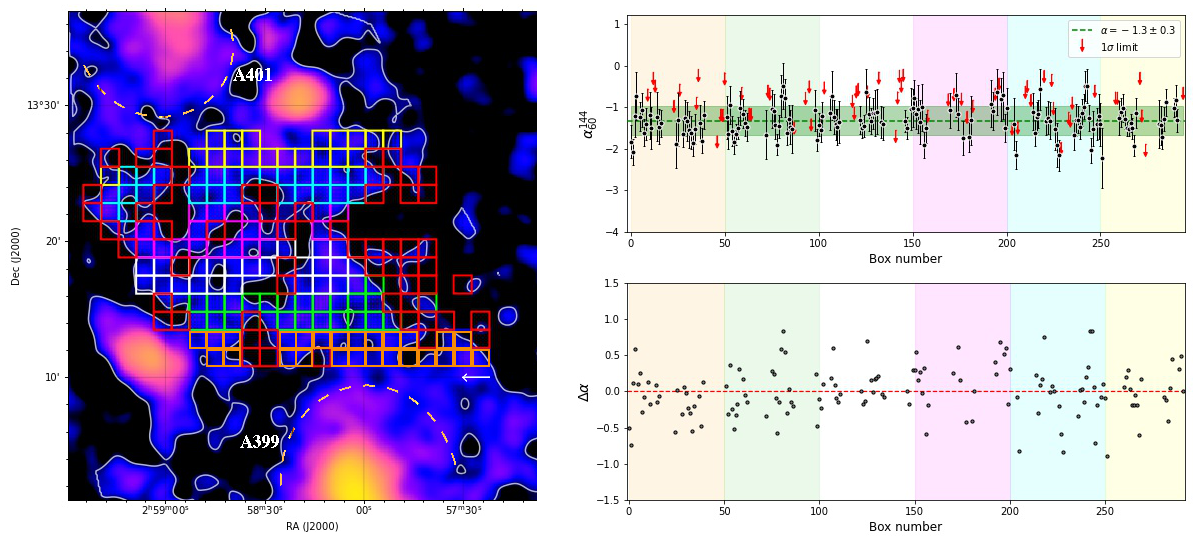}
      \caption{Spectral index distribution. \textit{Left panel:} LBA radio image at 60 MHz and resolution of $80''$ with overlaid $2\sigma_{\rm rms}$ contours and the grid to extract the spectral index between 60 and 144 MHz. Different areas of the bridge are covered by different colored cells. The colored cells covers the LBA emission above $2\sigma_{\rm rms}$ and the red cells covers the $1\sigma_{\rm rms}$ LBA emission and $2\sigma_{\rm rms}$ HBA emission. \textit{Top right:} Spectral index distribution across the bridge. The black points are values extracted from each cell, separated by color, and the red arrows are $1\sigma_{\rm rms}$ limits extracted from the red cells. The dashed green horizontal line is the mean spectral index, while the filled green horizontal region represents the standard deviation. \textit{Bottom right:} distribution of residuals ($\Delta\alpha$) of $\alpha$ with respect to the mean spectral index.}
         \label{fig:boxes}
\end{figure*} 
The radio spectrum of the bridge is shown in \Fig\ref{fig:spectrum} (top panel), the plotted values are listed in \T\ref{tab:flux_values}. The observations at 60 and 140~MHz and the upper limit derived at 400~MHz show a steepening towards increasing frequency. This feature, as well as the steep spectrum found between 144 and 400~MHz, is hard to reconcile with a shock origin scenario. However, spectral steeping is a key prediction of the turbulent re-acceleration model presented in \cite{brunetti2020}. In their work, they investigated several re-acceleration scenarios coupled with the result of a cosmological simulation, showing how the steepening changes as a function of magnetic field amplification efficiency ($\eta_{B})$ and particle acceleration times  ($\tau_{\rm acc}$). In particular they identified for their analysis a simulated system that resembles the A399-A401 bridge \citep{govoni2019, dominguez2019}. The initial spectrum of relativistic electrons is evolved solving the Fokker-Planck equations assuming a \textit{single-zone} model, i.e. assuming average quantities, such as thermal densities and magnetic fields, that are measured in each cell of the simulated bridge region at a fixed time. This is done for different values of particle acceleration times ($\tau_{\rm acc}$), from $\sim$10~Myr up to $\sim$10~Gyr (see \citealt{brunetti2020}, Fig. 2). 
The spectrum  shown in \Fig\ref{fig:spectrum} (bottom panel) is obtained from the spectrum presented in \cite{brunetti2020} (see Fig.~3, for $\eta_{B}\sim0.05$). To reproduce our observed data, we have to re-scale the spectrum from a surface of $\sim3.9$ Mpc$^{2}$ to $\sim2.2$ Mpc$^{2}$, which is the area covered by the LBA detection, excluding the radio halos. Additionally, it was necessary to exclude all cells with a generated $\tau_{\rm acc}<$ 200~Myr, which make up for $\sim$3\% of the volume of the simulated bridge. Therefore, the observed LOFAR luminosities and the uGMRT limit set a constraint on the particles acceleration times to values longer than 200~Myr, which are generated in the majority of the cells in the simulation.

Finally, we investigated how the spectral index value could vary in different areas of the bridge. The distribution of the spectral index is likely related to the contribution of turbulence and re-acceleration processes across the extended emission. We created a grid covering the emission inside the $2\sigma_{\rm rms}$ contours in the 60 MHz image (\Fig\ref{fig:boxes}, left panel).
Each grid cell is one beam size ($80''\times80''$). We extract the value of the spectral index between 60 and 144 MHz in each colored cell. Since we are computing the spectral index over the $2\sigma_{\rm rms}$ LBA detection, we are considering the emission component with steepest spectra and/or weakest surface brightness. To check this bias, in the red cell we extract the $1\sigma_{\rm rms}$ emission and evaluate a limit on the spectral index with the $2\sigma_{\rm rms}$ emission in the 144 MHz image. The distribution of the spectral index in the bridge is shown in \Fig\ref{fig:boxes} top right panel, and in the bottom right panel we show the distribution of the residuals, $\Delta\alpha$, between each value of spectral index extracted and the mean $\alpha$ value. The values of each cell are consistent around the mean value inside the standard deviation for most points. Despite a larger scatter being observed for some cells ($0.5<\Delta\alpha<0.9$), they do not appear to be spatially correlated. The red arrows represent the limits extracted from the red cells, generally flatter than the mean value. However, we also note the limits at the level of the measured spectral index, or steeper. The spectral index is consisted with being constant across the region, and there is no evidence for any systematic trends across the bridge region. \\

\section{Conclusions}

For the first time, we have determined the spectral index for the emission of a radio bridge, connecting the two pre-merging galaxy clusters A399 and A401. So far, the radio bridge was only detected at 144~MHz, therefore we analysed new LOFAR LBA data at 60~MHz to constrain the spectral index of the emission.

We measured an integrated spectral index for the bridge between 60 and 144 MHz of $\alpha_{60}^{144}= -1.44 \pm 0.16$. We also investigated the spectral index distribution, which gives insights into the contribution of turbulence and re-acceleration processes causing the extended emission. From the spectral index and associated errors maps, the distribution shows no systematic gradients in the bridge regions.

Combining the two LOFAR detection and the uGMRT limit found at 400~MHz, we are able to produce a comprehensive comparison of the obtained radio spectrum of the bridge with the synchrotron spectrum predicted by theoretical models. The steep spectral index derived between 144 and 400~MHz already challenged the shock-acceleration origin scenario \citep[as proposed in][]{govoni2019}. Moreover, we find that the steepening of the spectrum above 144, while hardly reconcilable with the shock acceleration scenario, it can be explained by the turbulent acceleration models investigated by \cite{brunetti2020}. Our observations allow us to constrain the particle acceleration time and, in turn, the volume-filling factor of the particle distribution in the turbulent re-acceleration model. Short acceleration times (corresponding to re-acceleration occurring in regions that occupy a small fraction of the bridge volume) generate shallower spectra, disfavoured by our observations. Conversely, large acceleration time ($\tau_{\rm acc}>$ 200~Myr) for particles that occupy most of the bridge volume,  are consistent with our data. The fact that the emission in the 60~MHz image appears less volume-filling than at 144~MHz, is likely related to the sensitivity limitations of the LBA observations.
The Fermi-II origin scenario suggested by these observations requires the presence of significant turbulent motions in most of the bridge volume. Moreover, the aforementioned scenario assumes the presence of a volume-filling reservoir of low energy electrons ($\gamma \leq 10^{3}$) whose existence requires further observational evidence, and also not quantitatively predicted by simulations yet. Finally, the $B\geq 0.3\, \mu$G magnetic field required by this model are large for such peripheral regions, and might be detected by the forthcoming generation of polarisation surveys \citep{heald2020}.

\begin{acknowledgements}
AB acknowledges financial support from the ERC Starting Grant `DRANOEL', number 714245. FdG acknowledges support from the ERC Consolidator Grant ULU 101086378. FV acknowledges the financial support from the Cariplo "BREAKTHRU" funds Rif: 2022-2088 CUP J33C22004310003. AB acknowledges financial support from the European Union - Next Generation EU. MB acknowledges funding by the Deutsche Forschungsgemeinschaft under Germany's Excellence Strategy -- EXC 2121 ``Quantum Universe'' --  390833306. RJvW acknowledges support from the ERC Starting Grant ClusterWeb 804208.

\end{acknowledgements}

\bibliographystyle{aa}
\bibliography{bibliography}

\begin{appendix}
\section{Injection procedure}\label{appendix}

As explained in \ref{models}, to incorporate the upper limit on the bridge emission at 400~MHz with uGMRT observations in the radio spectrum, it is necessary to perform the injection procedure \citep[see e.g.][]{venturi2008, bernardi2016, bonafede2017, duchesne2022, nunhokee2021} starting from the emission detected at 60~MHz. In this work, we extend the procedure presented in \cite{pignataro2023}, taking into accountspatial variations of the noise pattern in the uGMRT image.
This follows from a generalisation of the injection approach, where we do not only inject at the center of the image, (i.e. the location where the bridge is detected at lower frequencies), but also in different locations in the field. In fact, analysing the noise pattern in the final uGMRT image, over which the injection is performed, we find a non-uniform distribution.
Therefore, to place a more conservative limit on the flux density of the bridge, we perform the injection of visibilities three times: once at the known location of the bridge in the center of the image, once north-east from the center, and once south-west from the center. The final limit is the result of an average of the three injections.\\
The injection procedure can be summarised as follows (see \cite{pignataro2023} for an extensive description):
\begin{itemize}
    \item From the model image of the LOFAR detection at 60~MHz, we created a mask by including only the emission from the bridge.
    \smallskip
    \item The bridge model image is extrapolated to the uGMRT observation central frequency, with varying spectral index between $-4\leq\alpha\leq0$ with steps of $\Delta\alpha = 0.25$. The model images are additionally multiplied with the uGMRT primary beam model to take into account the attenuation of the primary beam in uGMRT observation.
    \smallskip
    \item The final bridge model images are then Fourier transformed into visibilities \citep[\texttt{WSClean -predict,}][]{offringa2014} that are injected into the uGMRT source-subtracted, calibrated visibilities. The prediction of the model visibilities in the original dataset takes into account the missing short baselines or flagged data of the uGMRT observation. We then deconvolved and imaged with the same parameters, uv sampling, and image and visibility weighting as the non-injected data images. In this way, we produced an $80^{\prime\prime}$ resolution mosaic injected image at the central frequency of 400~MHz for each spectral index, $\alpha$.
    \smallskip
    \item We define the ratio $R(\alpha)=S_{400}^{\rm inj}(\alpha)/S_{400}$ which measures how bright, given a certain spectral index value, $\alpha$, the injected bridge emission (i.e. the flux density $S_{400}^{\rm inj}(\alpha)$) is with respect to the image background ($S_{400}$). We computed the ratio $R(\alpha)$ each time we performed the injection in a different location. We then compute the average of the three ratios, $\langle R(\alpha)\rangle$.
    \smallskip
    \item We finally evaluate the cumulative probability distribution function of $\langle R(\alpha)\rangle$. We find that the spectral index of the bridge has a limit of $\alpha_{l}< -1.75 $ with a 95\% confidence level. This is different from the one presented in \citep{pignataro2023} because we started from a different injected model, and followed a more robust procedure.
\end{itemize}
The limit on the spectral index translates on a upper limit on the bridge integrated flux density, that is $S_{400}<60$~mJy.
\end{appendix}

\end{document}